\documentclass[runningheads]{llncs}
\usepackage{graphicx}
\usepackage{amsmath}
\usepackage{amssymb}
\usepackage{booktabs}
\usepackage{multirow}
\begin{document}
\title{Graph Attention Collaborative Similarity Embedding for Recommender System}
\titlerunning{GACSE}
%
\author{
Jinbo Song\inst{1} \and
Chao Chang\inst{2} \and
Fei Sun\inst{3} \and
Zhenyang Chen\inst{2} \and
Guoyong Hu\inst{2} \and
Peng Jiang\inst{2}
}
\authorrunning{Song et al.}
%
\institute{
The Institute of Computing Technology of the Chinese Academy of Sciences, China\\\email{songjinbo18s@ict.ac.cn} \and
Beijing Kuaishou Technology Co., Ltd., China\\\email{\{changchao, chenzhenyang, huguoyong, jiangpeng\}@kuaishou.com}\and
Alibaba Group, China\\\email{ofey.sf@alibaba-inc.com}
}
\maketitle              
\begin{abstract}
We present Graph Attention Collaborative Similarity Embedding (GACSE), a new recommendation framework that exploits collaborative information in the user-item bipartite graph for representation learning. Our framework consists of two parts: the first part is to learn explicit graph collaborative filtering information such as user-item association through embedding propagation with attention mechanism, and the second part is to learn implicit graph collaborative information such as user-user similarities and item-item similarities through auxiliary loss. We design a new loss function that combines BPR loss with adaptive margin and similarity loss for the similarities learning. Extensive experiments on three benchmarks show that our model is consistently better than the latest state-of-the-art models.

\end{abstract}

\section{Introduction}
Personalized recommendation plays a pivotal role in many internet scenarios, such as e-commerce, short video recommendations and advertising. Its core method is to analyze the user's potential preferences based on the user's historical behavior, to measure the possibility of the user to select a certain item and to tailor the recommendation results for the user. 

One of the major topics to be investigated in the personalized recommendation is Collaborative Filtering (CF) which generates recommendations by taking advantage of the collective wisdom from all users. Matrix factorization (MF)~\cite{fm} is one of the most popular CF model, which decomposes the interaction matrix between the user and item into discrete vectors and then calculates the inner product to predict the connected edges between the user and item. Neural Collaborative Filtering (NCF)~\cite{ncf} predict the future behavior of users by learning the historical interactions between users and items. It employs neural network instead of traditional matrix factorization to enhance the non-linearity of the model. In general, there are two key components in learnable CF models---1) embeddings that represent users and items by vectors, and 2) interaction modeling, which formulates historical interactions upon the embeddings. 

Despite their prevalence and effectiveness, we argue that these models are not sufficient to learn optimal embdeddings. The major limitation is that the embdeddings does not explicitly encode collaborative information propagated in user-item interaction graph. Following the idea of representation learning in graph embedding, Graph Neural Networks (GNN) are proposed to collect aggregate information from graph structure. Methods based on GraphSAGE~\cite{GCN} or GAT~\cite{GAT} have been applied to recommender systems. For example, NGCF~\cite{NGCF19} generates user and item embeddings based on the information propagation in the user-item bipartite graph; KGAT~\cite{KGAT19} adds a knowledge graph and entities attention based on the bipartite graph and uses entity information to more effectively model users and items. Inspired by the success of GNN in recommendation, we build a embedding propagation and aggregating architecture based on attention mechanism to learn the variable weight of each neighbor. The attention weight explicitly represents the relevance of interaction between user and item in bipartite graph.

Another limitation is that many existing model-based CF algorithms leverage only the user-item associations available in user-item bipartite graph. The effectiveness of these algorithms depends on the sparsity of the available user-item associations. Therefore, other types of collaborative relations, such as user-user similarity and item-item similarity, can also be considered for embedding learning. Some works~\cite{FactorizationItemEmbedding,Node2Vec} exploit higher-order proximity among users and items by taking random walks on the graph. A recent work~\cite{CSE} presents collaborative similarity embedding (CSE) to model direct and in-direct edges of user-item interactions. The effectiveness of these methods lies in sampling auxiliary information from graph to augment the data for representation learning. 

Based on the above limitation and inspiration, in this paper, we propose a unified representation learning framework, called Graph Attention Collaborative Similarity Embedding (GACSE). In the framework, the embedding is learned from direct, user-item association through embedding propagation with attention mechanism, and indirect, user-user similarities and item-item similarities through auxiliary loss, user-item similarities in bipartite graph. Meanwhile, we combine adaptive margin in BPR loss~\cite{rendle2012bayesian} and similarity loss to optimize GACSE. 

The contributions of this work are as follows:
\begin{itemize}
\item We propose GACSE, a graph based recommendation framework that combines both attention propagation \& aggregation in graph and similarity embedding learning process. 
\item To optimize GACSE, we introduce a new loss function, which, to the best of our knowledge, is the first time to combine both BPR loss with adaptive margin and similarity loss for similarity embedding learning.  
\item We compare our model with state-of-the-art methods and demonstrate the effectiveness of our model through quantitative analysis on three benchmark datasets.
\item We conduct a comprehensive ablation study to analyze the contributions of key components in our proposed model.

\end{itemize}

\section{Related Work}
In this section, we will briefly review several lines of works closely related to ours, including general recommendation and graph embedding-based recommendation.
\subsection{General Recommendation}
Recommender systems typically use Collaborative Filtering (CF) to model users’ preferences based on their interaction histories~\cite{ASurveyofCollaborativeFilteringTechniques,AdvancesinCollaborativeFiltering}. Among the various CF methods, item-based neighborhood methods~\cite{itemcf} estimate a user’s preference on an item via measuring its similarities with the items in her/his interaction history using a item-to-item similarity matrix. User-based neighborhood methods find similar users to the current user using a user-to-user similarity matrix, following by recommending the items in her/his similar users' interaction history. Matrix Factorization (MF)~\cite{fm,pmf} is another most popular one, which projects users and items into a shared vector space and estimate a user’s preference on an item by the inner product between user's and items' vectors. BPR-MF~\cite{rendle2012bayesian} optimizes the matrix factorization with implicit feedback using a pairwise ranking loss. Recently, deep learning techniques has been revolutionizing the recommender systems dramatically. One line of deep learning based model seeks to take the place of conventional matrix factorization~\cite{autoRec,CDAR,ncf}. For example, Neural Collaborative Filtering (NCF) estimates user preferences via Multi-Layer Perceptions (MLP) instead of inner product.

\subsection{Graph based Recommendation}
Another line is to integrate the distributed representations learning from user-item interaction graph. GC-MC~\cite{GC-MC} employs a graph convolution auto-encoder on user-item graph to solve the matrix completion task. HOP-Rec~\cite{HOP} employs label propagation and random walks on interaction graph to compute similarity scores for user-item pairs. NGCF~\cite{ncf} explicitly encodes the collaborative information of high-order relations by embedding propagation in user-item interaction graph. PinSage~\cite{PinSage} utilizes efficient random walks and graph convolutions to generate embeddings which incorporate both graph structure as well as node feature information. Multi-GCCF~\cite{Multi-GCCF} constructs two separate user-user and item-item graphs. It employs a multi-graph encoding layer to integrate the information provided by the user-item, user-user and item-item graphs.

\section{Our Model}
\begin{figure*}[ht]
\centering
\includegraphics[width=\linewidth]{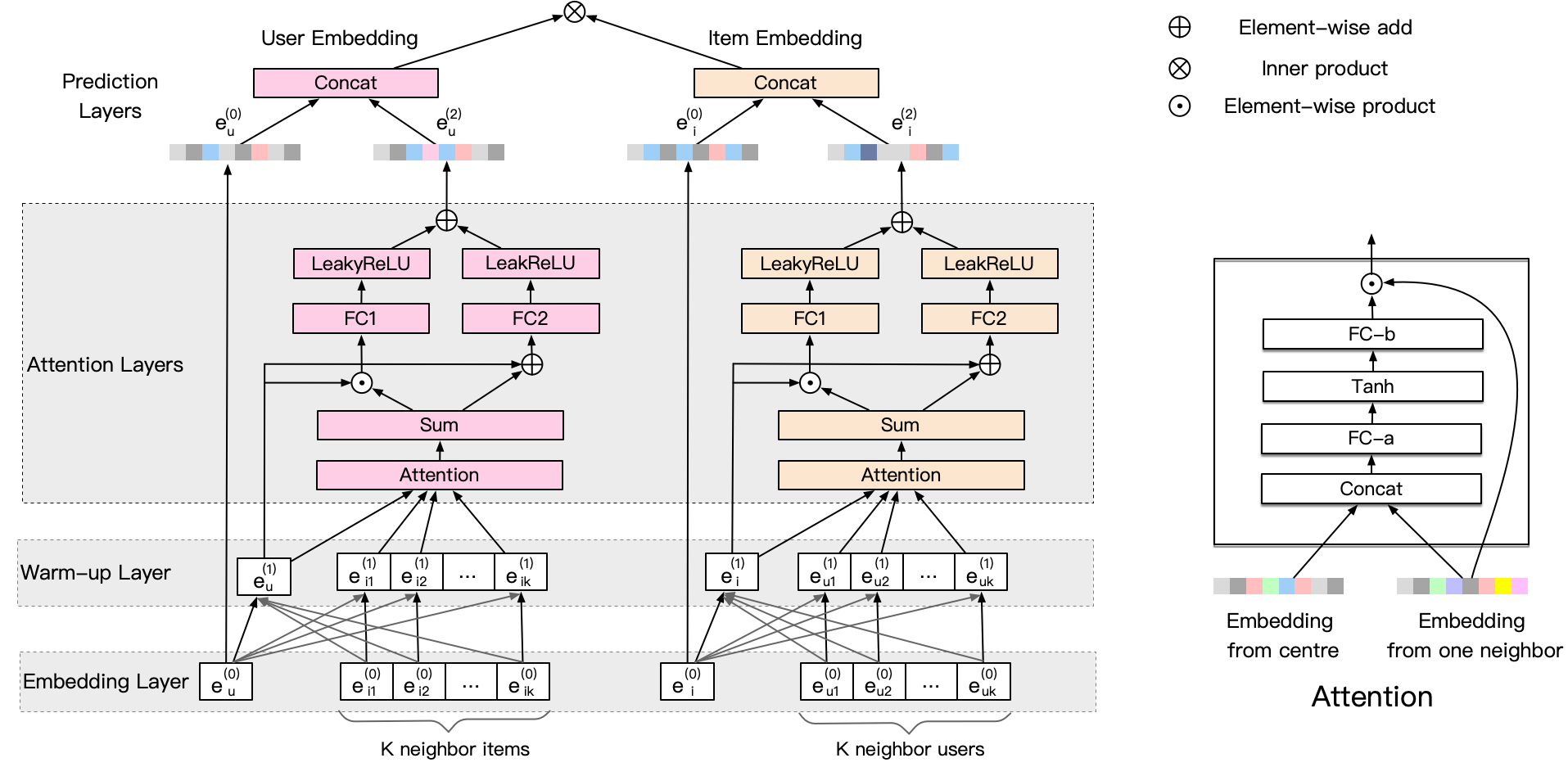}
\caption{An illustration of GACSE model architecture. The flow of embedding is presented by the arrowed lines. \text{FC1} and \text{FC2} are shared parameters on user embedding side and item embedding side. An illustration of attention aggregation is shown on the left. $\text{FC-a}$ transforms the concatenated vector into a new vector. $\text{FC-b}$ transforms vector into attention score.}
\label{fig:gasce}
\end{figure*}

In this section, we introduce our proposed model called GACSE. The overall framework is illustrated in~Figure~\ref{fig:gasce}. There are three components in the model: (1) an embedding layer that can map users and items from one hot vector to initial embeddings; (2) an embedding propagation layer, which consists of two sub-layers: a warm-up layer that propagates and aggregates graph embeddings with equal weight, and an attention layer that uses attention mechanism to perform non-equal weight aggregation on the embedding of neighboring nodes; and (3) a prediction layer that concatenates embeddings from embedding layer and attention layer, then outputs affinity score between user and item. The following descriptions take user as central node, if there is no special instruction, it is also applicable to item as centre node.

\subsection{Embedding Layer}
Embedding layer aims at mapping the ids of user $u$ and item $i$ into embedding vectors $\mathbf{e}_u^{(0)} \in \mathbb{R}^d$ and $\mathbf{e}_i^{(0)} \in \mathbb{R}^d$, where $d$ denotes the embedding dim. We use a trainable embedding lookup table to build our embedding layer for embedding propagation:
\begin{equation}
    \mathbf{E} = [ \overbrace{\mathbf{e}_{u_1}^{(0)},\cdots,\mathbf{e}_{u_N}^{(0)}}^{\text{users}\,\text{embeddings}}\; , \; \overbrace{\mathbf{e}_{i_1}^{(0)},\cdots,\mathbf{e}_{i_M}^{(0)}}^{\text{items}\,\text{embeddings}} ]
\end{equation}
where $N$ is number of users and $M$ is number of items. 

\subsection{Embedding Propagation Layer}

In order to establish the embedding propagation architecture for collaborative information in graph, we define a embedding propagation layer. Our embedding propagation layer consists of two parts: (1) a warm-up layer and (2) an attention layer. Both layers have two steps: embedding propagation and aggregation.



\subsubsection{Warm-up Layer}
To make the model expand the receptive field and grasp the structure of the user-item bipartite graph, we set up a warm-up layer based on the GNN~\cite{GCN,Hamilton2017InductiveRL} embedding propagation architecture. 

\paragraph{Warm-up Propagation}In the warm-up layer, we set all the embeddings to have equal weight. We define warm-up embedding propagation function as:
\begin{equation}
    \left\{
        \begin{aligned}
            &\mathbf{m}_{i\rightarrow u}^{(0)}=\pi^{(0)}_{(u,i)}\mathbf{W}_{0}\mathbf{e}_{i}^{(0)} \\
            &\mathbf{m}_{u\rightarrow u}^{(0)}=\mathbf{W}_{0}\mathbf{e}_{u}^{(0)}
        \end{aligned}
    \right.
\end{equation}
where $\mathbf{W}_0 \in \mathbb{R}^{d_{1}\times d_{0}}$ is the trainable weight matrix to distill important information in embedding propagation. $d_{0}$ is the dimension of $\mathbf{e}^{(0)}$, and $d_{1}$ is the dimension of transformation.
$\mathbf{m}_{i\rightarrow u}^{(0)}$ is the embedding from item $i$ to user $u$. $\mathbf{m}_{u\rightarrow u}^{(0)}$ is self-connection of $u$. $\pi_{(u,i)}^{(0)}$ is the weight of the embedding that $i$ passes to $u$.

In embedding propagation of warm-up layer, we set the propagation weight $\pi^{0}_{(u,i)}$ to be equal. Inspired by GCN~\cite{GCN}, we define weight of each user's interacted item as:

\begin{equation}
    \pi^{(0)}_{(u,i)}=\frac{1}{\sqrt{|\mathcal{N}_{u}\Vert\mathcal{N}_{i}|}}
\end{equation}
where $\mathcal{N}_{u}$ and $\mathcal{N}_{i}$ denote the first hop neighboring nodes of user u and item i.

\paragraph{Warm-up Aggregation} After receiving the embeddings from neighbor nodes, we need to aggregate these embeddings. We define embeddings aggregation function in warm-up layer as:
\begin{equation}
    \mathbf{e}^{(1)}_{u} = \sigma\bigl(\mathbf{m}_{u\rightarrow u}^{(0)} + \sum_{i\in \mathcal{N}_u}\mathbf{m}_{i\rightarrow u}^{(0)}\bigr)
\end{equation}
where $\sigma$ is nonlinear function such as $\text{LeakyReLU}$. Analogously, we can obtain item $i$'s embedding $\mathbf{e}^{(1)}_{i}$.

\subsubsection{Attention Layer}
Next, in order to further encode the variable weight of neighbors, we build an embedding propagation and aggregation architecture based on the attention mechanism. The attention mechanism explicitly captures the relevance of interaction between user and item in bipartite graph.

\paragraph{Attention Propagation}
Intuitively, the importance of each item that interacts with the user should be different. We introduce attention mechanism into embedding passing function:
\begin{equation}
    \left\{
        \begin{aligned}
            &\mathbf{m}_{i\rightarrow u}^{(1)} = \pi^{(1)}_{(u,i)} \mathbf{e}_{i}^{(1)} \\
            &\mathbf{m}_{u\rightarrow u}^{(1)}=\mathbf{e}_{u}^{(1)}
        \end{aligned}
    \right.
\end{equation}
where $\pi^{(1)}_{(u,i)}$ is attention weight.

Inspired by several kinds of attention score functions, we define score function of our model:
\begin{equation}
    \text{score}(\mathbf{e}_u^{(1)},\mathbf{e}_i^{(1)}) = \mathbf{V}^{\top}\text{tanh}\Bigl(\mathbf{P}\left[\mathbf{e}_{u}^{(1)}\Vert\mathbf{e}_{i}^{(1)}\right]\Bigr)
\end{equation}
where $\mathbf{V} \in \mathbb{R}^{d_2\times 1}$ and $\mathbf{P}\in \mathbb{R}^{d_2\times 2d_1}$ are trainable parameters. $d_2$ is the dimension of attention transformation. $\Vert$ denotes concatenate operation. After calculating attention score, we normalize it to get the attention weight via softmax function: 
\begin{equation}
    \pi^{(1)}_{(u,i)} = \frac{\text{exp}(\text{score}(\mathbf{e}^{(1)}_u,\mathbf{e}^{(1)}_i))}{\sum_{j\in \mathcal{S}_u}\text{exp}(\text{score}(\mathbf{e}^{(1)}_u,\mathbf{e}^{(1)}_j))}
\end{equation}
where $\mathcal{S}_u$ is a set of user $u$'s one hop neighboring items sampled in this mini-batch. 

\paragraph{Attention Aggregation}After attention massage passing, the attention aggregation function is defined as:
\begin{equation}
\begin{aligned}
    \mathbf{e}^{(2)}_u =& \sigma(\mathbf{W}_1(\mathbf{m}^{(1)}_{u\rightarrow u}+\sum_{i \in \mathcal{S}_u}\mathbf{m}^{(1)}_{i\rightarrow u})) +\\ & \sigma(\mathbf{W}_2(\mathbf{m}^{(1)}_{u\rightarrow u}\odot \sum_{i \in \mathcal{S}_u}\mathbf{m}^{(1)}_{i\rightarrow u}))
\end{aligned}
\end{equation}
where $\sigma$ is LeakyReLU non-linear function. $\mathbf{W}_1$, $\mathbf{W}_2 \in \mathbb{R}^{d_3\times d_1}$ are trainable parameters. $d_3$ is the dimension of attention aggregation. $\odot$ denotes element-wise product. Similar to NGCF, the aggregated embedding $\mathbf{e}^{(2)}_u$ does not only related to 
$\mathbf{e}^{(1)}_i$, but also encodes the interaction between $\mathbf{e}^{(1)}_u$ and $\mathbf{e}^{(1)}_i$. The interaction information can be represented by the element-wise product between $\mathbf{m}_{u\rightarrow u}$ and $\sum_{i \in \mathcal{S}_u}\mathbf{m}_{i\rightarrow u}$. Analogously, item $i$'s attention layer embedding $\mathbf{e}^{(2)}_i$ can be obtained. We incorporate the attention mechanism to learn variable weight  $\pi^{(1)}_{(u,i)}$ for each neighbor's propagated embedding $\mathbf{m}^{(1)}_{i\rightarrow u}$.

\subsection{Model Prediction}
After embedding passing and aggregation with attention mechanism, we obtained two different representations $\mathbf{e}_u^{(0)}$ and $\mathbf{e}_u^{(2)}$ of user node $u$; also analogous to item node $i$, we obtained $\mathbf{e}_i^{(0)}$ and $\mathbf{e}_i^{(2)}$. We choose to concatenate the two embeddings as follows:
\begin{align}
    \mathbf{e}_u^{*} = \mathbf{e}_u^{(0)} \Vert\, \mathbf{e}_u^{(2)},\quad\mathbf{e}_i^{*} = \mathbf{e}_i^{(0)} \Vert\, \mathbf{e}_i^{(2)}
\end{align}
where $\Vert$ denotes the concatenate operation. In this way, we could predict the matching score between user and item by inner product:
\begin{equation}
    y_{ui} = {\mathbf{e}_u^{*}}^{\top} \mathbf{e}_i^{*}
\end{equation}
More broadly, we can define the matching score between any two nodes a and b:
\begin{equation}
    y_{ab} = {\mathbf{e}_a^{*}}^{\top} \mathbf{e}_b^{*}
\end{equation}

\section{Optimization}
To optimize the GACSE model, we carefully designed our loss function. Our loss function consists of two basic parts: BPR loss with adaptive margin and similarity loss.

\subsection{BPR loss with adaptive margin}
We employ BRP loss for optimization, which considers the relative order between observed and unobserved interactions. In order to improve the model's discrimination of similar positive and negative samples, we define BPR loss with adaptive margin as:
\begin{equation}
\begin{aligned}
    \mathcal{L}_{\text{BPR}} = \frac{1}{|\mathcal{B}|}\sum_{(u,i,j)\in \mathcal{B}}-\sigma(y_{ui}-y_{uj} - \max(0,y_{ij}))
\end{aligned}
\end{equation}
where $\mathcal{B} \subseteq \{(u,i,j)|(u,i)\in \mathcal{R}^{+}, (u,j)\in \mathcal{R}^{-}\}$ denotes the sampled data of mini-batch. $\mathcal{R}^{+}$ denotes observed interactions, and $\mathcal{R}^{-}$ is unobserved interactions. $\sigma$ is softplus function. $\max(0,y_{ij})$ indicates that the more similar the positive and negative samples of a node are, the larger the margin of the loss function is.

\subsection{Similarity loss}

Other types of collaborative relations, such as user-user similarity and item-item similarity in graph, can also be considered for embedding learning. The introduce of similarity loss for both user-user and item-item pair can reduce the sparsity problem by augmenting the data for representation learning. In this paper, the 2-order neighborhood proximity of a pair of users (or items) is defined as the similarity.

In order to avoid similarity loss affecting the embedding in the embedding propagation, we only calculate between $\mathbf{E}$ and context mapping embedding matrices $\mathbf{E}^{\text{UC}}$ and $\mathbf{E}^{\text{IC}}$ for users and items, respectively. Context mapping embedding matrices are defined:
\begin{equation}
    \begin{aligned}
        \mathbf{E}^{\text{UC}} &= [\mathbf{e}^{\text{UC}}_{u_1},\cdots,\mathbf{e}^{\text{UC}}_{u_N}] \\
        \mathbf{E}^{\text{IC}} &= [\mathbf{e}^{\text{IC}}_{i_1},\cdots,\mathbf{e}^{\text{IC}}_{i_M}]
    \end{aligned}
\end{equation}
It should be noted that the dimensions of embeddings in $\mathbf{E}$, $\mathbf{E}^{\text{UC}}$ and $\mathbf{E}^{\text{IC}}$ are equal. The similarity loss for $\mathbf{e}^{(0)}$ with context embeddings $\mathbf{e}^{\text{UC}} \in \mathbf{E}^{\text{UC}}$ and $\mathbf{e}^{\text{IC}} \in \mathbf{E}^{\text{IC}}$ is defined as:

\begin{equation}
\begin{aligned}
     \mathcal{L}_{\text{similarity}} = &-\sum_{}\text{log}(\sigma(\mathbf{e}^{(0)\text{T}}_{u}\mathbf{e}^{\text{UC}}_{\text{u-pos}}))
     +\sum_{}\text{log}(\sigma(\mathbf{e}^{(0)\text{T}}_{u}\mathbf{e}^{\text{UC}}_{\text{u-neg}}))\\
     &-\sum_{}\text{log}(\sigma(\mathbf{e}^{(0)\text{T}}_{i}\mathbf{e}^{\text{IC}}_{\text{i-pos}}))
     +\sum_{}\text{log}(\sigma(\mathbf{e}^{(0)\text{T}}_{i}\mathbf{e}^{\text{IC}}_{\text{i-neg}}))
\end{aligned}
\end{equation}
where $\sigma$ is sigmoid function. $\mathbf{e}^{\text{UC}}_{\text{u-pos}}$ and $\mathbf{e}^{\text{UC}}_{\text{u-neg}}$ are a positive and negative samples of user $u$, respectively.  $\mathbf{e}^{\text{IC}}_{\text{i-pos}}$ and $\mathbf{e}^{\text{IC}}_{\text{i-neg}}$ are a positive and negative samples of item $i$, respectively. We employ random walk and negative sampling to construct positive and negative sample pairs for similarity loss.
\subsection{Overall Loss Function}
Finally, we get the overall loss function:
\begin{equation}
    \mathcal{L} = \mathcal{L}_{\text{BPR}} + \lambda_1\mathcal{L}_{\text{similarity}} + \lambda_2\Vert\Theta\Vert_{2}^{2}
\end{equation}
where $\Theta=\{\mathbf{E}, \mathbf{E}^{\text{UC}}, \mathbf{E}^{\text{IC}}, \mathbf{W}_0, \mathbf{W}_1, \mathbf{W}_2, \mathbf{V}, \mathbf{P}\}$. $\lambda_1$ controls the strength of BPR loss and $\lambda_2$ controls the $L_2$ regularization strength to prevent overfitting. We use mini-batch Adam~\cite{Adam} to optimize the model and update the parameters of model.

\section{EXPERIMENTS}
\subsection{Datasets}
We evaluate the proposed model on three real-world representative datasets: Gowalla\footnote{
https://snap.stanford.edu/data/loc-gowalla.htm}, Yelp2018\footnote{https://www.yelp.com/dataset/challeng} and Amazon-book\footnote{http://jmcauley.ucsd.edu/data/amazon/}. These datasets vary significantly in domains and sparsity. The statistics of the datasets are summarized in Table~\ref{tab:dataset}.
 
For each dataset, the training set is constructed by $80\%$ of the historical interactions of each user, and the remaining as the test set. We randomly select $10\%$ of interactions as a validation set from the training set to tune hyper-parameters. We employ negative sampling strategy to produce one negative item that the user did not act before and treat observed user-item interaction as a positive instance. To ensure the quality of the datasets, we use the 10-core setting, i.e., retaining users and items with at least ten interactions.

\begin{table}
\centering
\begin{tabular}{lrrr}  
\toprule
 &Gowalla&Yelp2018&Amazon-Book\cr
\midrule
\#Users & 29,858 & 45,919 & 52,643 \cr
\#Items & 40,981 & 45,538 & 91,599 \cr
\#Interactions & 1.027m & 1.185m & 2.984m\cr
Density & 0.084\% & 0.056\% & 0.062\% \cr
\bottomrule
\end{tabular}
\caption{Statistics of the datasets}
\label{tab:dataset}
\end{table}

\subsection{Experimental Settings}

To evaluate the effectiveness of top-K task in recommender system, we adopted Recall@K and NDCG@K, which has been widely used in~\cite{ncf,NGCF19}. In this paper, 1) we set K = 20; 2) all items that the user has not interacted with are the negative items; 3) all items is scored by each method in descend order except the positive ones used in the training set. Average metrics for all users in the test set is used for evaluation.

To verify the effectiveness of our approach, we compare it  with the following baselines: 
\begin{itemize}
\item \textbf{BPR-MF}~\cite{rendle2012bayesian} optimizes the matrix factorization with implicit feedback using a pairwise ranking loss.  

\item \textbf{NCF}~\cite{ncf} learns user's and item's embeddings from user–item interactions in a matrix factorization, which by a MLP instead of the inner product .

\item \textbf{PinSage}~\cite{PinSage} combines efficient random walks and graph convolutions to generate embeddings of nodes that incorporate both graph structure as well as node feature information.

\item \textbf{GC-MC}~\cite{GC-MC} is a graph auto-encoder framework based on differentiable embedding passing on the bipartite interaction graph. The auto-encoder produces latent user and item representations, and they are used to reconstruct the rating links through a bilinear decoder.

\item \textbf{NGCF}~\cite{NGCF19} explicitly encodes the collaborative signal of high-order relations by embedding propagation in user-item inter-action graph.

\item \textbf{Multi-GCCF}~\cite{Multi-GCCF} constructs two separate user-user and item-item graphs. It employs a multi-graph encoding layer to integrate the information provided by the user-item, user-user and item-item graphs.

\end{itemize}
We implement GACSE\footnote{For reproducibility, we share the source code of GACSE online:
https://github.com/GACSE/GACSE.git} with TensorFlow. 
The embedding size is fixed to 64 for all models. All models are optimized with the Adam optimizer, where the batch size is fixed at 1024. 
The learning rate of our model was set to 0.0001; $\lambda_1$ was set to $1\times10^{-4}$; $\lambda_2$ was set to $1\times10^{-5}$; Number of sampling neighbors was set to 64. Number of positive and  negative samples for similarity loss was set to 5. All hyper-parameters of the above baselines are either followed the suggestion from the methods' author or turned on the validation sets. We report the results of each baseline under its optimal hyper-parameter settings.

\subsection{Performance Comparison}
\renewcommand\tabcolsep{5.0pt}
\begin{table}
\centering
\begin{tabular}{lcccccc}  
\toprule
&\multicolumn{2}{c}{Gowalla}&\multicolumn{2}{c}{Yelp2018}&\multicolumn{2}{c}{Amazon-Book} \\ \cmidrule(lr){2-3} \cmidrule(lr){4-5} \cmidrule(lr){6-7}
&Recall&NDCG&Recall&NDCG&Recall&NDCG\cr
\midrule
BPR-MF           & 0.1291 & 0.1878 & 0.0494 & 0.0662 & 0.0250 & 0.0518 \cr
NCF              & 0.1326 & 0.1985 & 0.0513 & 0.0719 & 0.0253 & 0.0535 \cr
PinSage          & 0.1380 & 0.1947 & 0.0612 & 0.0750 & 0.0283 & 0.0545 \cr
GC-MC            & 0.1395 & 0.1960 & 0.0597 & 0.0741 & 0.0288 & 0.0551 \cr
NGCF             & 0.1547 & \underline{0.2237} & 0.0581 & 0.0719 & 0.0344 & 0.0630 \cr
Multi-GCCF       & \underline{0.1595} & 0.2126 & \underline{0.0667} & \underline{0.0810} & \underline{0.0363} & \underline{0.0656} \cr
\textbf{GACSE}   & \textbf{0.1654} & \textbf{0.2328} & \textbf{0.0672} & \textbf{0.0836} & \textbf{0.0386} & \textbf{0.0703} \cr
\midrule
\%Improv.        & 3.70\% & 4.06\% & 0.75\% & 3.21\% & 6.34\% & 7.16\% \cr
\bottomrule
\end{tabular}
\caption{Overview performance comparison. Bold scores are the best in each column, while underlined scores are the second best. Improvements are statistically significant.}
\label{PerformanceComparison}
\end{table}

\subsubsection{Overall Comparison}
Table~\ref{PerformanceComparison} summarized the best results of all models on three benchmark dataset. The last row is the improvements of GACSE relative to the best baseline.

BPR-MF method gives the worst performance on all datasets since the inner product cannot capture complex collaborative signals. NCF outperforms BPR-MF on all datasets consistently. Compared with BPR-MF, the main improvement of NCF is that MLP can model the nonlinear feature interactions between user and item embeddings.

Among all the baseline methods, graph based methods (e.g., PinStage, GC-MC, NGCF, Multi-GCCF) consistently outperform general methods (e.g., BPR-MF, NCF) on all datasets. The main improvement is that graph based model explicitly models the graph structure in embedding learning.

Multi-GCCF are the strongest baseline. It outperforms other baselines on all datasets except NDCG on Gowalla. NGCF  gives the best performance of NDCG on Gowalla. They all employ embedding propagation to obtain neighbor's information and stack multiple embedding propagation layers to explore the high-order connectivity. This verifies the importance of capturing collaborative signal in the embedding function. Moreover, Multi-GCCF compared three different multi-grained representations fusion methods.

According to the results, GACSE preforms best among all baselines on three datasets in terms of all evaluation metrics. It improves over the best baseline method by $3.70\%$, $0.75\%$, $6.34\%$ in terms of Recall on Gowalla, Yelp2018 and Amazon-book. It gains $4.06\%$, $3.21\%$, $7.16\%$ NDCG improvements against the best baseline on Gowalla, Yelp2018 and Amazon-book respectively.
Compared with Multi-GCCF and NGCF, GACSE builds an embedding propagation and aggregation architecture based on the attention mechanism. The attention mechanism enable GACSE to  learn variable weights of embedding propagation for neighbors explicitly. Meanwhile it obtains high-order implicit collaborative information between user-user and item-item through similarity loss. 

\subsubsection{Ablation Analysis}
\begin{table}
\centering
\begin{tabular}{l cc cc cc}  
\toprule
&\multicolumn{2}{c}{Gowalla}&\multicolumn{2}{c}{Yelp2018} \\ \cmidrule(lr){2-3} \cmidrule(lr){4-5} \cmidrule(lr){6-7}
&Recall&NDCG&Recall&NDCG\cr
\midrule
GACSE               & 0.1654 & 0.2328 & 0.0672 & 0.0836 \cr
\hline
\multirow{2}*{GACSE-sl}  & 0.1632 & 0.2334 & 0.0641 & 0.0805 \cr
                   & (-1.33\%) &  (+0.26\%) & (-4.61\%) &  (-3.71\%) \cr
\hline
\multirow{2}*{GACSE-am}       & 0.1468 & 0.2149 & 0.0571 & 0.0728 \cr
                   & (-11.25\%) & (-7.68\%) & (-15.03\%) & (-12.92\%) \cr
\bottomrule
\end{tabular}
\caption{Ablation studies of GACSE. GACSE-sl means GACSE without similarity loss. GACSE-am means GACSE without adaptive margin.}
\label{AblationAnalysis}
\end{table}


Table~\ref{AblationAnalysis} reports the influences of similarity loss and adaptive margin of GACSE on Gowalla and Yelp2018 datasets. As expected, the performance degrades greatly after removing adaptive margin and similarity loss. This confirms the importance of adaptive margin and similarity loss for embedding learning. 
Adaptive margin can improve model's discrimination for positive and negative sample with similar embeddings. Similarity loss for both user-user and item-item pair can reduce the sparsity problem by augmenting the data for representation learning. Similarity loss and adaptive margin can enhance the effectiveness of attention mechanism for embedding propagation and aggregation.

\subsubsection{Test Performance w.r.t. Epoch}




\begin{figure}[ht]
\centering
\begin{minipage}[t]{0.48\textwidth}
\centering
\includegraphics[width=\linewidth]{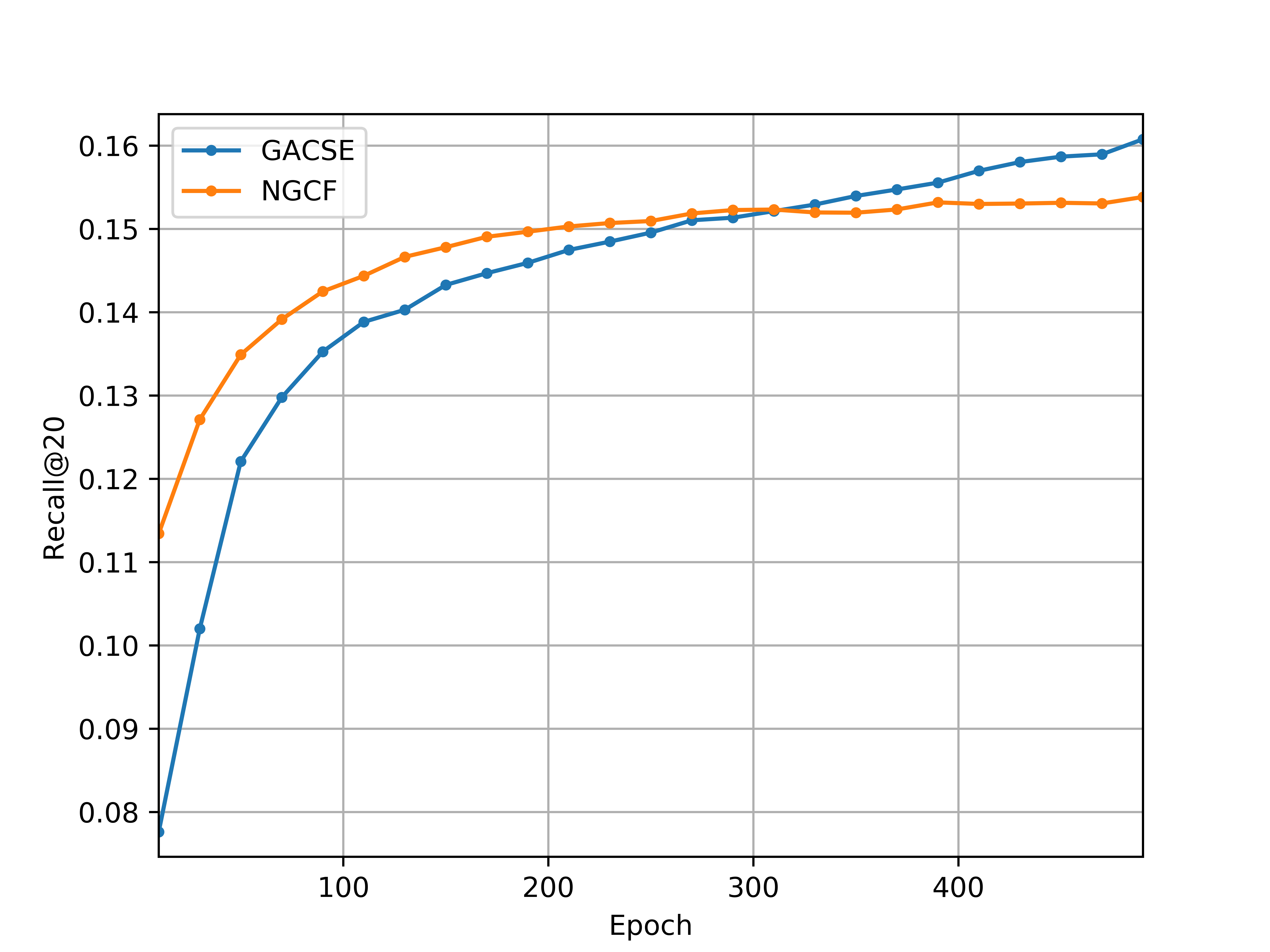}
\caption{Recall@20 on Gowalla}
\label{fig:recall_gowalla}
\end{minipage}
\begin{minipage}[t]{0.48\textwidth}
\centering
\includegraphics[width=\linewidth]{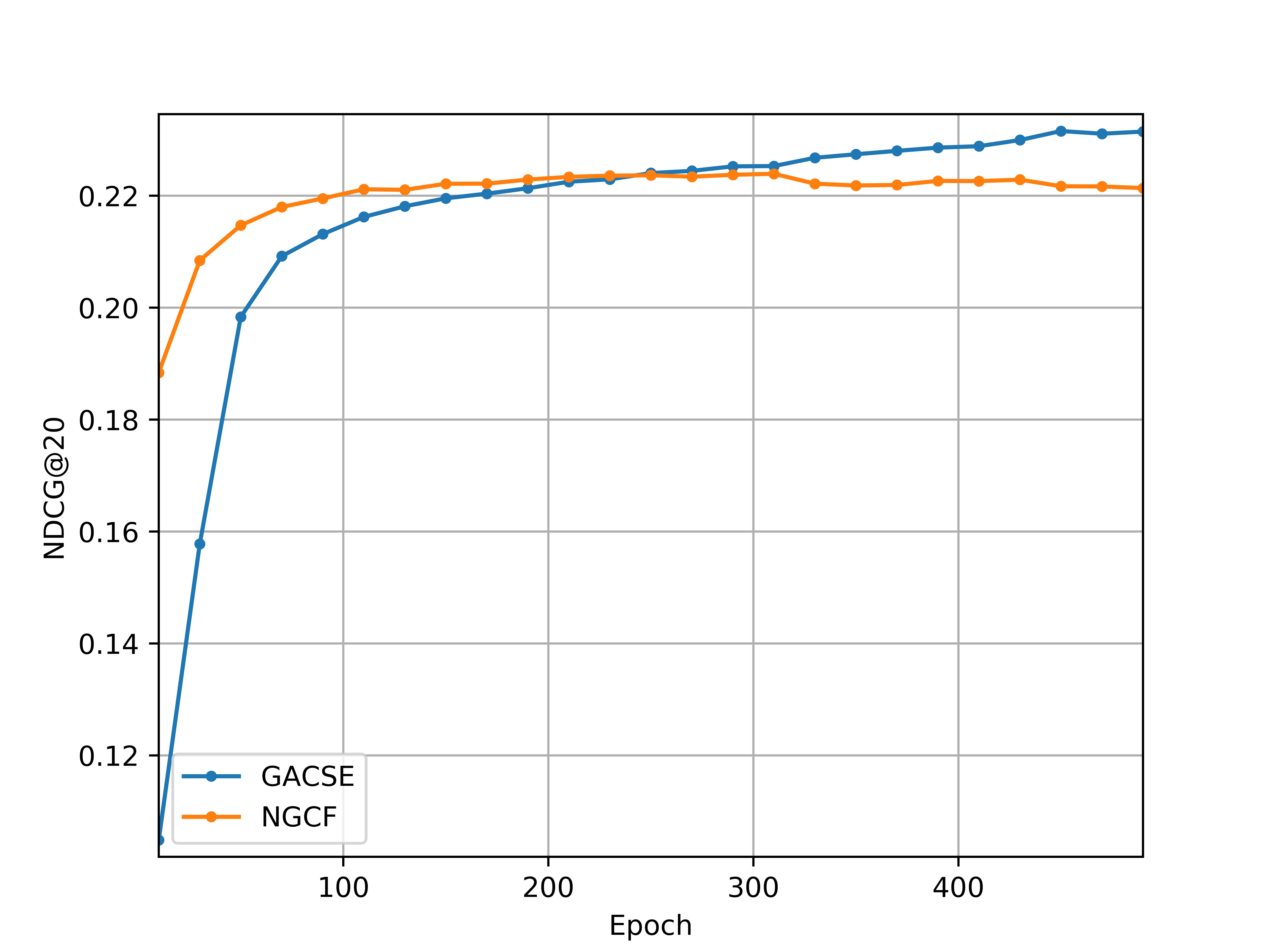}
\caption{NDCG@20 on Gowalla}
\label{fig:ndcg_gowalla}
\end{minipage}
\end{figure}

\begin{figure}[ht]
\centering
\begin{minipage}[t]{0.48\textwidth}
\centering
\includegraphics[width=\linewidth]{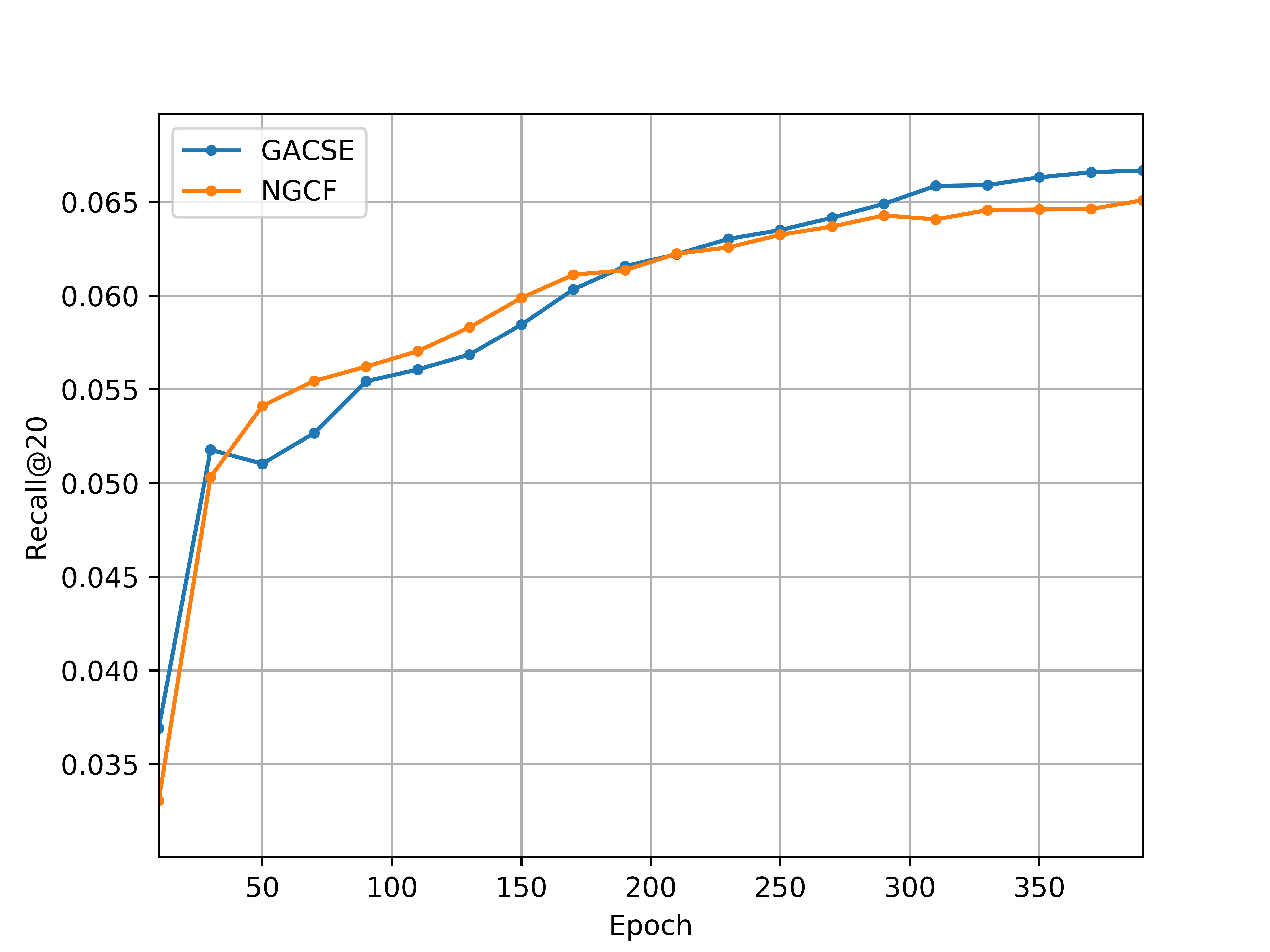}
\caption{Recall@20 on Yelp}
\label{fig:recall_yelp}
\end{minipage}
\begin{minipage}[t]{0.48\textwidth}
\centering
\includegraphics[width=\linewidth]{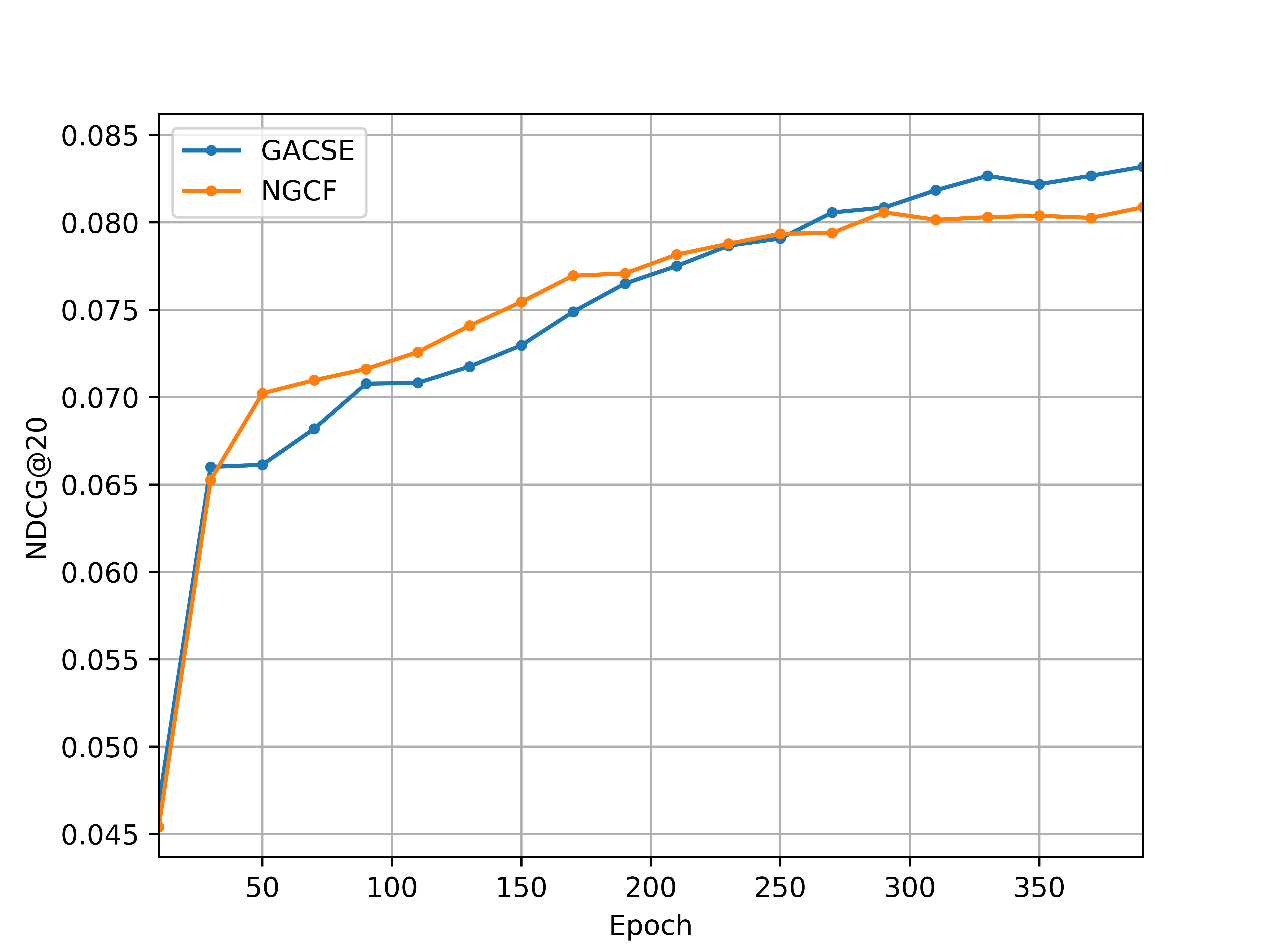}
\caption{NDCG@20 on Yelp}
\label{fig:ndcg_yelp}
\end{minipage}
\end{figure}

Figure 7 shows the test performance w.r.t. recall of each epoch of MF and NGCF. Due to the space limitation, we omit the performance w.r.t. ndcg which has the similar trend. We can see that, NGCF exhibits fast convergence than MF on three datasets. It is reasonable since indirectly connected users and items are involved when optimizing the interaction pairs in mini-batch. Such an observation demonstrates the better model capacity of NGCF and the effectiveness of performing embedding propagation in the embedding space.


\section{CONCLUSION AND FUTURE WORK}

In this work, we explicitly incorporated collaborative signal and indirect similarities into the embedding function. We proposed a unified representation learning framework GACSE, in which the embedding is learned from direct user-item interaction through attention propagation, and indirect user-user similarities and  item-item similarities through auxiliary loss, user-item similarities in bipartite graph. In addition, we combine adaptive margin in BPR loss and similarity loss to optimize GACSE. Extensive experimental results on three real-world datasets show that our model outperforms state-of-the-art baselines.

Several directions remain to be explored. A valuable direction is to incorporate rich side information into GACSE instead of just modeling user \& item ids. 
Another interesting direction for the future work would be exploring multi-task \& multi-object embedding learning on heterogeneous graph for recommender system.

\bibliographystyle{ACM-Reference-Format}
\bibliography{sample-base}

\vspace{12pt}

\end{document}